\documentclass[twocolumn,showpacs,preprintnumbers,amsmath,amssymb]{revtex4}

\usepackage{amsbsy}
\usepackage{graphicx}

\begin{document}

\title{Magnetic Field Distribution Due To Domain Walls \\ In Unconventional Superconductors}

\author{N.~A.~Logoboy}

\affiliation{Racah Institute of Physics, The Hebrew University of
Jerusalem, Jerusalem 91904, Israel}

\date{\today}

\begin{abstract}

Steady-state properties of 180$^{0}$ Bloch domain wall (DW) in
superconducting ferromagnet (SCFM) are studied. The distribution of
magnetic field above and below the surface of the SCFM due to the
permanent magnetization supercurrent flowing in the DW plane is
calculated by solving Maxwell equations supplemented by London
equation. It is shown that part of the magnetic flux of the two
neighboring domains closures in the nearest vicinity of the surface
of the sample giving rise to declination of the line of the force
from being parallel to the DW plane. As a result, the value of the
normal component of magnetic field at the surface of the sample
reaches only half of the value of the bulk magnetic flux. At the
distances of the order of value of London penetration depth the
magnetic field decreases as inverse power law due to the long-range
character of dipole-dipole interaction. The last two circumstances
are important for comparison the calculated magnetic field with the
data obtained by the methods on measurement of normal component of
magnetic field, e. g. Hall probe technique, aimed to confirm the
existence of magnetic order parameter in unconventional
superconductor.
\end{abstract}


\maketitle

\section{\label{sec:Introduction}Introduction}

Unconventional superconductors are a subject of intensive
experimental and theoretical studies during past decade
\cite{Felner}-\cite{Sonin}. Coexistence of superconductivity and
magnetism results in a number of unusual phenomena, which have both
fundamental and application interests. Unconventional
superconductors, such as $\textrm{Sr}_{2}\textrm{RuO}_{4}$
\cite{Luke}, $\textrm{ZrZn}_{2}$ \cite{Pfleiderer} and
$\textrm{UGe}_{2}$ \cite{Saxena}, which are characterized by the
state with broken time-reversal symmetry (TRSB), possess the
magnetic structure related to non-zero orbital magnetic moment due
to spin-triplet $p$-wave state of multi-component SC order
parameter. The properties of such superconductors are not trivial.
In particular, the macroscopic magnetization is not well-defined
quality, and the equations of motion cannot be expressed in terms of
local magnetization \cite{Braude1}.

Starting from the pioneering work \cite{Volovik}, \cite{Sigrist1} on
investigation of non-uniform states of superconducting order
parameter (see, also review \cite{Sigrist2}), the planar defects
such as domain walls (DW) and surfaces have been attracted recently
a lot of interest \cite{Bjornsson}, \cite{Kealey}, \cite{Tamegai},
\cite{Dolocan} due to new experiment possibilities, e. g. scanning
superconducting quantum interference device (SQUID) and Hall probe
technique, on investigation of magnetic field distribution resulting
from spontaneously generated superconducting currents which can
serve as a prove of TRSB origin of superconducting order parameter
in unconventional chiral superconductors.

Although, the magnetostatic fields are screened by superconducting
current, metastable domain walls (DWs), as topologically stable
planar defects, may exist even in the Meissner state
\cite{Sonin_Felner}. Thus, the domain structure of SCFM cannot be
ignored. The anomalies in the local magnetization loop for
Sr$_{2}$RuO$_{4}$ near $B=0$ are considered as a strong indication
of the presence of chiral domains \cite{Tamegai}. The strong
coordinate dependence of magnetization of such a $2$D magnetic
defect creates the intrinsic magnetic field which interacts with
superconducting (SC) current. For unconventional superconductors the
discontinuity of magnetic induction at the DW can be interpreted as
an effective magnetization, and the contribution of DW current to
the energy density can be transformed to an effective Zeeman term
providing the possibility of excitation of orbital magnetization
waves by incident electromagnetic field \cite{Braude1}.

In present publication we solve the Maxwell equations for
distribution of magnetic field inside the sample and above it due to
the presence of planar defect, such as a DW, show rapid decrease of
the magnetic field in the vicinity of sample surface, discuss the
long-range origin of this field resulting in existence of the tails
at the distance $\sim \lambda$ (London penetration depth) and
compare our results with recent experimental data on measurements of
the magnetic field in Sr$_{2}$RuO$_{4}$ by SQUID \cite{Bjornsson}
and Hall probe technique \cite{Kirtley}. It is shown that for
infinitely thin DW the stray fields above the sample depend only the
magnetic field jump at the DW, but not on the details of DW
structure.

\section{\label{sec:Basic Equations}Basic Equations}

Let us consider the $180^0$ DW of Bloch type in a semi-infinite
sample of superconducting ferromagnet occupying $y\le 0$. We assume,
that the surface of a crystal is parallel to the $x-z$ plane at
$y=0$, and the domain wall, being parallel to the $y-z$ plane,
separates two domains with magnetization $M(x)$ along the $+y$ or
$-y$ direction provided that the quality parameter
$\alpha=H_{K}/4\pi M_{0}>1$, e.g. the field of magnetic
crystallographic anisotropy of the easy-axis ( $y$-axis ) type
($H_{K}$) is high enough in order to stabilize the system against
the flip of the domain magnetization into the $x$-$z$ plane. We
restrict ourselves to the simplest case when the London penetration
length $\lambda$ exceeds the DW thickness $\Delta$. This means that
at the spatial scales of the order of $\Delta$, the domain wall
structure is governed by the exchange and magnetic crystallographic
energies and is not affected by the kinetic energy associated with
superconducting currents. Thus, we assumed that the next hierarchy
of characteristic lengths for SCFM is fulfilled: $\Delta<\lambda <
D$, where $D$ is the size of the domain. We consider the
steady-state properties of DW deep in the bulk.

\begin{figure}
  \includegraphics[width=0.4\textwidth]{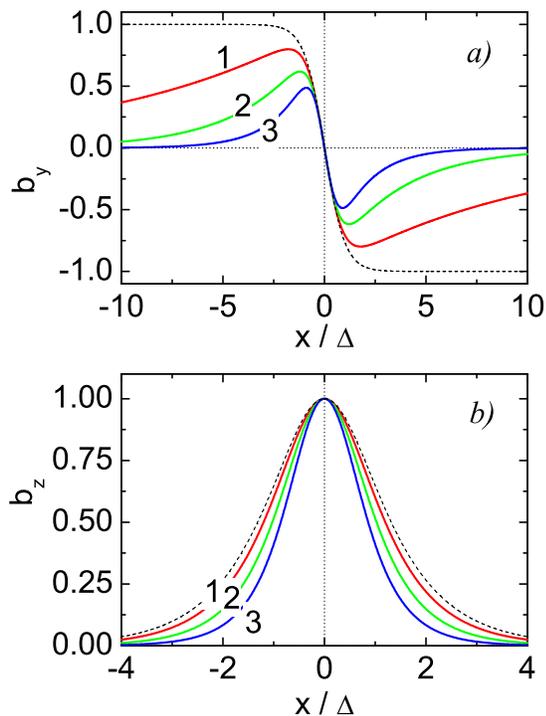}
\caption{\ Shown are the $y$-component $(a)$ and $z$-component $(b)$
of magnetic induction for Bloch DW at constant value of the domain
wall width $\Delta$ and variable London penetration depth $\lambda$.
The indexes 1, 2 and 3 correspond to $\delta=\Delta/\lambda=0.1$,
$0.3$ and $0.6$ respectively. The dashed line shows the
$y$-component $(a)$ and $z$-component $(b)$ of magnetization which
are the limiting values for corresponding components of magnetic
induction at $\lambda \longrightarrow \infty$, e.g. in neglecting of
screening by superconducting current.} \label{Induction}
\end{figure}

The free energy of the superconducting ferromagnet can be written as
follows \cite{Sonin}:

\begin{eqnarray} \label{eq:free energy}
   F&=&\int d^3x
   \left\{%
    2\pi\alpha \left[ M^{2}_\perp+\Delta^2(\partial_{i} M_{i})(\partial_{i} M_{i})\right]+\frac{1}{8\pi\lambda^{2}}\right. \nonumber\\
&&\times\left.\left(\frac{\Phi_{0}}{2\pi}\mathbf{\nabla}\phi-\mathbf{A}\right)^{2}+\frac{B^{2}}{8\pi}-\mathbf{B}\cdot\mathbf{M}
\right\},\quad%
\end{eqnarray}
where $\phi$ is the phase of the superconducting order parameter,
$\Phi_{0}=\hbar c /2e$ is the magnetic flux quantum, $\mathbf{M}$ is
the magnetization, and $M_\perp$ is the magnetization component
perpendicular to the easy axis $\hat{\mathbf{z}}$, which is the
normal to the sample surface. The vector potential $\mathbf{A}$
determines the magnetic induction $\mathbf{B}=\mathbf
{curl}~\mathbf{A}$. The first and second terms in Eq.~(\ref{eq:free
energy}) describe the magnetic crystallographic and exchange
energies correspondingly, the third term is kinetic energy of
superconducting current, and the last two terms relate to
magnetostatic energy. The length $\Delta$ characterizes the
stiffness of the spin system and is also the DW width. The
magnetization can be expressed in polar coordinates, $ \mathbf{M} =
M_{0}(\sin\phi\ \sin\theta,\cos\theta,\cos\phi\ \sin\theta)$ with
$M_{0}=g\mu_{B}s/a^{3}$ the saturation magnetization and $a$ the
lattice constant.

\begin{figure}[b]
  \includegraphics[width=0.4\textwidth]{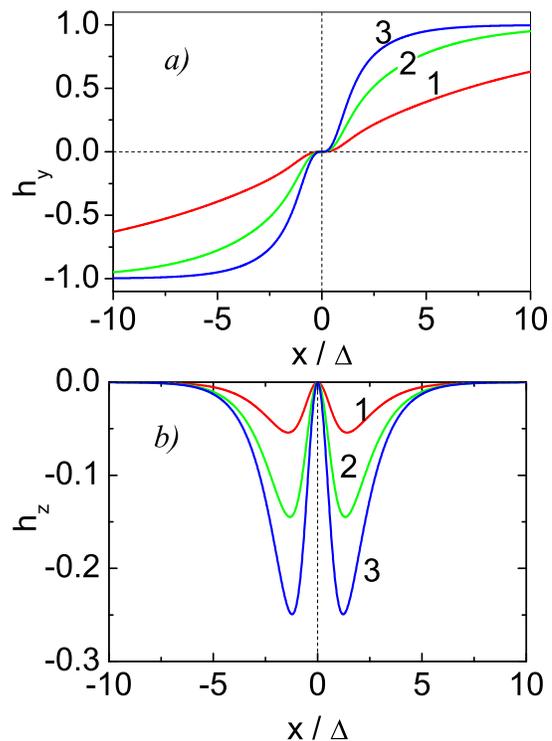}
\caption{\ Shown are the $y$-component $(a)$ and $z$-component $(b)$
of magnetic field for Bloch DW at constant value of the domain wall
width $\Delta$ and variable London penetration depth $\lambda$. The
indexes 1, 2 and 3 correspond to $\delta=\Delta/\lambda=0.1$, $0.3$
and $0.6$ consequently.} \label{Magnetic field}
\end{figure}

\begin{figure}[t]
  \includegraphics[width=0.4\textwidth]{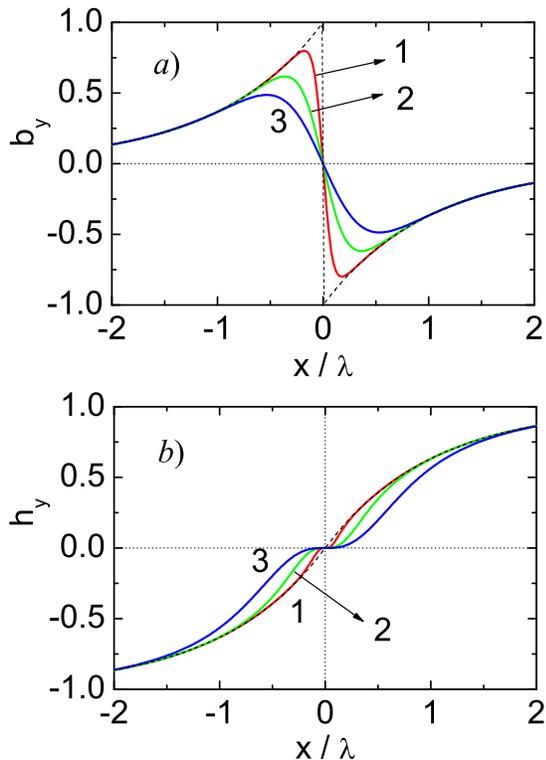}
\caption{\ Shown are the $y$-components of magnetic induction ($a$)
and magnetic field ($b$) for Bloch DW at constant value of the
London penetration depth $\lambda$ and variable DW width $\Delta$.
The indexes 1, 2 and 3 correspond to $\delta=\Delta/\lambda=0.1$,
$0.3$ and $0.6$ consequently. The dash lines correspond the limit
values for magnetic induction ($a$) and magnetic field ($b$) in case
of infinitely thin DW \cite{Sonin}.} \label{limit}
\end{figure}

\subsection{\label{sec:GS}Ground state of $180^{0}$ Bloch domain wall}

The ground state of the SCFM is defined by:
\begin{equation} \label{eq:derivatives}
\frac{\delta F}{\delta q_{i}}=0, \\
\end{equation}
where $q_{i}=\theta,\phi,\mathbf{A}$. Minimization of the free
energy density Eq.~(\ref{eq:free energy}) with respect to azimuthal,
$\phi$, polar, $\theta$, angles and vector potential $\mathbf{A}$
leads to the system of non-linear differential equations describing
the ground state of SCFM with DW of Bloch type. In case of
$\delta=\Delta/\lambda <1$ the solution of these equations are as
follows:
\begin{eqnarray}
\theta_{0}=2\tan^{-1} e^{x/\Delta},\quad \phi_{0}=0, \qquad\qquad
\nonumber\\
b^{(0)}_{y}=\cos\theta_{0}e^{(x/\lambda) \cos\theta_{0}},\quad
b^{(0)}_{z}=\sin\theta_{0}e^{(x/\lambda) \cos\theta_{0}}
 \label{eq:ground state},
\end{eqnarray}
where we used the reduced values for the components of magnetic
induction, $b^{(0)}_{y}=B^{(0)}_{y}/4\pi M_{0}$ and
$b^{(0)}_{z}=B^{(0)}_{z}/4\pi M_{0}$. The condition $\delta <1$
allows us to neglect the influence of Meissner current on the
structure of the DW, which is defined entirely by the magnetic
anisotropy and exchange energy. The screening action of the current
results in decreasing of magnetic induction at the distance of $\sim
\lambda$ from the center of the domain wall $x=0$. The graphical
representation of the solutions (\ref{eq:ground state}) are shown in
Fig.~\ref{Induction} for different values of London penetration
depth $\lambda$ at constant DW width $\Delta$. The distribution of
magnetic field $\mathbf{H}=\mathbf{B}-4\pi \mathbf{M}$, created by
Meissner current, are shown in Fig.~\ref{Magnetic field}. It follows
from Fig.~\ref{Induction} and Fig.~\ref{Magnetic field} that the
screening effect of superconducting current degreases with
increasing of London penetration depth $\lambda$.

At constant $\lambda$ in limit case when the DW width $\Delta
\longrightarrow 0$, the jump of the tangential component of
magnetization $M_{y}$ at the plane of the geometric domain boundary
$y-z $ defines the current sheet responsible for the jump of
tangential component of magnetic induction $B_{y}$ \cite{Sonin}. The
distributions of tangential components of magnetic induction $B_{y}$
and magnetic field $H_{y}$ of Meissner current at different values
of the DW width $\Delta$ and constant London penetration depth
$\lambda$ are shown in Fig.~\ref{limit}. The results of our
calculation based on Eq.~(\ref{eq:ground state}) show the decrease
of maximum of magnetic induction splitting with increase of DW width
$\Delta$ (see, Fig.~\ref{limit}$a$) which confirms the effect of
smoothing due to finite DW width \cite{Bluhm}.

It follows from Maxwell equation $\mathbf
{curl}~\mathbf{H}=\mathbf{j}$ that there exists a two-component
screening current $\mathbf j(x) =(0,j_{y}(x),j_{z}(x))$, which is
the result of 2-dimensional structure of the DW Eq.~(\ref{eq:ground
state}).

\section{\label{sec:Results and Discussion}Results and Discussion}

To calculate the distribution of magnetic field near the surface of
SCFM we neglect the domain wall $\Delta$, assuming that
$\Delta<<\lambda$. This assumption does not affect the results
qualitatively, but essentially simplifies the problem. In the end we
shall discuss the effects of finite DW width $\Delta \ne 0$.

\subsection{\label{sec:Infinitely Thin DW}Infinitely thin DW ($\Delta \to 0$)}

The difference between a normal FM and a SCFM is important at
distances larger than $\Delta$: while in normal FMs the magnetic
field $\mathbf{H}=\mathbf{B}-4\pi\mathbf{M}$ vanishes and the
magnetic induction $\mathbf{B}= 4\pi\mathbf{M}$ is constant inside
domains, in SCFMs the magnetic induction $\mathbf{B}$ is confined in
the Meisssner layers of width $\lambda$ \cite{Sonin}:
\begin{equation} \label{eq:Magnetic Induction}
B_{y}= \pm \frac{1}{2}\delta B_{y} e^{\pm x/\lambda}~,
 \end{equation}
where the upper and the lower signs correspond to $x<0$ and $x>0$
respectively, and $\delta B_{y}$ is the magnetic induction
splitting. Thus the Meissner currents $j_{z}=-(c/4\pi)
\partial_{x}B_{y}$ screen out the main bulk of domains from the magnetic
induction. This screened magnetic induction Eq.~(\ref{eq:Magnetic
Induction}) influences at the distribution of magnetic field above
the sample surface and, in principle, can be detected by Hall
probes. To find the distribution of magnetic field above the surface
of the sample due to magnetic induction Eq.~(\ref{eq:Magnetic
Induction}) the standard procedure of solving Maxwell equation in
magnetic media ($y<0$) and vacuum ($y>0$) with appropriate boundary
conditions at the surface located at $y=0$, e. g. the continuity of
normal component of magnetic induction and tangential component of
magnetic field, is used (see, e. g. \cite{Chikazumi}). The results
of calculation can be represented in explicit form:
\begin{eqnarray} \label{eq:Vacuum}
 h_{x}=-\frac{2}{\pi}\int_{0}^{+\infty} dk~\frac{}{}\frac{k\exp{(-ky)}}{\widetilde {k}(k+\widetilde{k})}\cos{(kx)}~,
  \nonumber \\
 h_{y}=\frac{2}{\pi}\int_{0}^{+\infty} dk~\frac{}{}\frac{k\exp{(-ky)}}{\widetilde
 {k}(k+\widetilde{k})}\sin{(kx)}
\end{eqnarray}
for the components of the reduced magnetic field
$h_{i}=2H_{i}/\delta B_{i}$ in vacuum, and
\begin{eqnarray} \label{eq:Sample}
 b_{x}=-\frac{2}{\pi}\int_{0}^{+\infty} dk~\frac{k\exp{(\widetilde{k}y)}}{\widetilde{k}(k+\widetilde{k})}\cos{(kx)}~,
  \nonumber \\
 b_{y}=\frac{2}{\pi}\int_{0}^{+\infty}
 dk~\frac{k[\widetilde{k}+k(1-\exp{(\widetilde{k}y))}]}{\widetilde{k}^{2}(k+\widetilde{k})}\sin{(kx)}
\end{eqnarray}
for the components of the reduced magnetic induction
$b_{i}=2B_{i}/\delta B_{i}$ in the sample. In
Eqs.~(\ref{eq:Vacuum}),(\ref{eq:Sample}) we used the notation
$\widetilde{k}=(k^{2}+\lambda^{-2})^{1/2}$. The expression for
$h_{y}$ (last equation in (\ref{eq:Vacuum})) in slightly different
form was calculated in \cite{Bluhm}.

\begin{figure}[t]
  \includegraphics[width=0.4\textwidth]{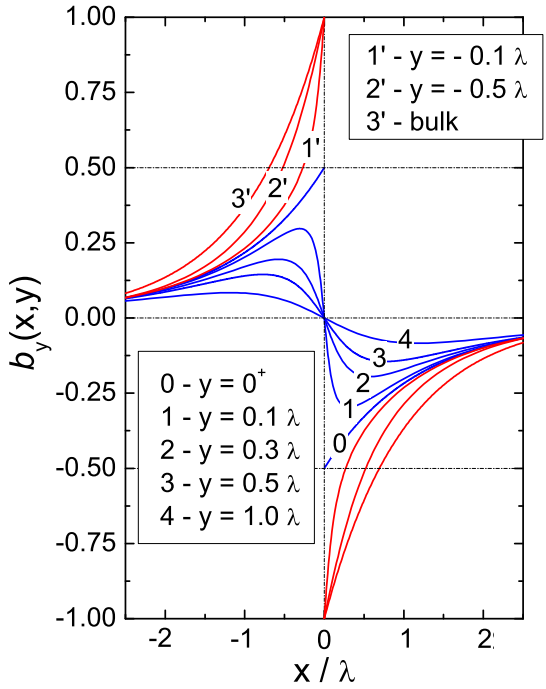}
\caption{\ (color online) Shown are the $y$-component of magnetic
field in vacuum above the surface of SCFM (blue line) and in the
sample (red line) for one Bloch DW lying in $y-z$ plane. The
presence of the identical tails in the distribution at large
distances are due to long-range origin of dipole-dipole interaction
in accordance to Eq.~(\ref{eq:Distribution}).} \label{Normal
Component}
\end{figure}

The components of magnetic field in vacuum Eqs.~(\ref{eq:Vacuum})
can be calculated by introducing the complex potential
$\psi=\psi(w)$ which is analytical function of complex variable
$w=y-ix$. Thus, the complex magnetic field
\begin{equation} \label{eq:Complex Magnetic Field}
h(w)=h_{y}(x,y)+ih_{x}(x,y)
\end{equation}
is derived from $\psi(w)$ by $h(w)=-\partial_{w} \psi(w)$. The
complex potential $\psi(w)$ can be expressed through special
functions
\begin{equation} \label{eq:Complex Potential}
\psi(w)=-i\frac{2}{\pi}\frac{1}{w}-i\partial_{w}[\mathbf{H}_{0}(w)-N_{0}(w)],
\end{equation}
where $\mathbf{H}_{0}(w)$ and $N_{0}(w)$ are zero-order Struve and
Neumann functions correspondingly (see, e. g. \cite{Abramowitz}).
For $\mid w \mid \ge \lambda$, the difference
$\mathbf{H}_{0}(w)-N_{0}(w) \approx 2/\pi w$, therefore, the main
contribution to the potential $\psi (w)$ is due to the first term in
Eq.~(\ref{eq:Complex Potential}) which allows to calculate the
asymptotic distribution of the component of magnetic field above the
sample ($y\ge \lambda$)
\begin{equation} \label{eq:Distribution}
h_{x}=\frac{2}{\pi}\lambda^{2}\frac{x^{2}-y^{2}}
{(x^{2}+y^{2})^{2}}, \quad
h_{y}=\frac{4}{\pi}\lambda^{2}\frac{xy}{(x^{2}+y^{2})^{2}}.
\end{equation}
In particular, it follows from Eq.~(\ref{eq:Distribution}), that at
$\mid w \mid
>> 1$ the value of magnetic field decreases as inverse power law, e. g. $\mid h \mid \sim \mid w \mid^{-2}$, due to long-range magnetostatic
interaction.

The results of numerical calculations of normal to the sample
surface component of magnetic field in vacuum above the surface of
the sample $h_{y}$ Eq.~(\ref{eq:Vacuum}) and in the sample $b_{y}$
Eq.~(\ref{eq:Sample}) are represented in Fig.~\ref{Normal Component}
which shows that at the sample surface $y=$0, magnetic induction
splitting equals half of it bulk value $\delta B_{y}$ and are
characterizes by the rapid decrease with the distance above the DW.
Thus, at the distance of $y=$0.1$\lambda$ above the DW the magnetic
field decays till the third of the bulk value $\delta B_{y}$. At $y
\sim \lambda$, the magnetic field splitting reaches about 0.05 of it
bulk value and decays with $x$ as inverse power low
Eq.~(\ref{eq:Distribution}) due to the long-range origin of
dipole-dipole interaction.

In a limit of infinitely thin DW, $\Delta \to 0$, the stray fields
in vacuum above the superconducting ferromagnet depend only on the
jump of magnetic induction at DW plane, but not on the details of DW
structure. In this case, the tangential to the sample surface
component of magnetic field is non-analytic function of $y$ and is
characterized by the discontinuous jump at the DW position, $y=0$.
In next subsection we take into account the effects of small
($\Delta<<\lambda$), but finite DW width $\Delta \ne 0$.

\subsection{\label{sec:GS}Finite DW width ($\Delta \ne 0$)}

To consider the effects related to the finite DW width, $\Delta\ne
0$, we assume the linear distribution of magnetic induction in the
DW deep in the bulk, e.g. $B_{y}\approx(1/2)\delta B_{y}(x/\Delta)$
at $\mid x \mid \le \Delta$, which is not affected by
superconducting currents, inasmuch as $\Delta<<\lambda$. This
assumption significantly simplifies the calculation and allows to
express the results in analytic form. It can be shown that for
finite DW width the kernels in Eqs.~(\ref{eq:Vacuum}) and
(\ref{eq:Sample}) are modified by the multiplier
$\sin(k\Delta)/k\Delta$. It results in slight suppression of
$y$-component of magnetic field and restores the analyticity of
$x$-component of magnetic field. The discontinuous jump of
tangential component of magnetic field at the position of DW is
replaced by the value $h^{max}_{x}=(1/\pi)\ln(\gamma \delta)$, where
$\gamma \approx 1.78107$ is Euler-Mascheroni constant.

Recent experiments on imaging of magnetic field distribution above
the surface of the sample of Sr$_{2}$RuO$_{4}$ by scanning SQUID and
Hall probe microscopy has revealed no evidence for existence of DWs
in this unconventional superconductor \cite{Kirtley}. This negative
result can be understood in the framework of the developed theory.
If the DWs exist, the maximum value of magnetic induction due to
permanent current flowing in the plane of the DW does not exceed the
lower critical field, e. g. $\delta B_{y}/2\sim H_{c1}\approx$30G
above which the domain structure is unstable due to formation of
Abrikosov vortices. Thus, at the distance of $y=\lambda \approx$190
nm, the normal component of magnetic field due to the presence of DW
is of order of 3G which definitely improbable by the used methods.

\acknowledgments

The stimulated inspiring discussions with Prof. E. B. Sonin are
highly appreciated. This work has been supported by the grant of the
Israel Academy of Sciences and Humanities.

  \end{document}